\documentclass[11pt]{article}
\usepackage{DGfest,epsfig}
\usepackage{times}

\usepackage{lscape}                                           %
\usepackage[centertags]{amsmath}
\usepackage{amsfonts} \usepackage{amssymb}\usepackage{eufrak}
\usepackage{graphicx}
\usepackage{a4}
\usepackage{lscape}

\def\be{\begin{equation}}
\def\ee{\end{equation}}
\def\bea{\begin{eqnarray}}
\def\eea{\end{eqnarray}}

\begin{document}
\baselineskip 11.5pt
\title{BOUNDARY STATE FOR MAGNETIZED D9 BRANES AND ONE-LOOP CALCULATION}

\author{PAOLO DI VECCHIA }

\address{NORDITA, Blegdamsvej 17, DK-2100 Copenhagen \O, Denmark}

\author{ANTONELLA LICCARDO, RAFFAELE MAROTTA and FRANCO PEZZELLA}

\address{Dipartimento di Scienze Fisiche, Universit\`{a} di Napoli and
  INFN, Sezione di Napoli, \\
Complesso Universitario Monte S. Angelo, ed. G  - via Cintia -
I-80126 Napoli, Italy}
\author{IGOR PESANDO}
\address{Dipartimento di Fisica Teorica, Universita' di Torino \\
and INFN Sezione di Torino, Via P. Giuria 1, I-10125 Torino, Italy}

\maketitle\abstracts{
We construct the boundary state describing magnetized D9 branes in
$R^{3,1} \times T^6$ and we use it to compute the  annulus and
M{\"{o}}bius amplitudes. We derive from them, by using open/closed string duality, the number of Landau levels on the torus $T^d$.}

\section{Introduction}
\label{intro}

String theories are perfectly consistent theories in ten
non-compact dimensions. However, in order for them to be consistent
with particle phenomenology, six of the ten dimensions have to be
compactified. The compactification procedure produces a number of
fields called moduli whose
vacuum expectation value cannot be fixed in perturbation theory if
supersymmetry is preserved. Therefore a major problem that one has to solve
in string theory  is to find ways to fix the vacuum expectation values of
those moduli in order to be able to compare with particle
phenomenology.

Recently, magnetized D9 branes have been used to produce semirealistic
models where the moduli are stabilized  and the tadpoles
are canceled out~\footnote{ See for instance Ref.s ~\cite{AM,AKM} and
  Ref.s therein.}. This has been done by mostly using the Born-Infeld
action that encodes their properties.

In this paper we give a string description of magnetized D9 branes
 by determining
the corresponding boundary state in $R^{3,1} \times T^6$
following closely the procedure outlined in
Ref.~\cite{IGOR}. We use it to provide a straightforward derivation of
the one-loop amplitudes corresponding to the annulus and M{\"{o}}bius
diagrams and
generalizing the results of Ref.~\cite{BT2} to the case of an
arbitrary NS-NS $B_2$-field and arbitrary wrappings.

In the next section we will construct the boundary state corresponding to
a number of magnetized D9 branes and in the third section we will use it
for computing one-loop amplitudes.

\section{The boundary state for  magnetized D9 branes in $R^{3,1} \times T^6$}
\label{boundary}

In order to determine the boundary state of magnetized D9 branes on
$R^{3,1} \times T^6$ we start from the open string channel by
considering the action which describes the interaction of a string with
in general two arbitrary abelian gauge fields $ A^{(0)}, A^{(\pi)}$ acting
respectively at the endpoints of the string $\sigma =0, \pi$. In the
following we
will consider only the six compact directions omitting the four
non-compact ones because the boundary state corresponding to them
has been already determined (see the two reviews on the
boundary state~\cite{ANTOI,ANTOII}).
We will also omit to discuss the part of the boundary
state corresponding to the world-sheet fermion degrees of freedom that
can be found in Ref.s~\cite{ANTOI,ANTOII}.

The action of an open string in a closed
toroidal string background interacting with two arbitrary abelian gauge fields
with constant field strength is given by~\footnote{Hereafter we
  closely follow Ref.~\cite{GO}}:
\begin{eqnarray}
S = - \frac{1}{4 \pi
  \alpha'}
\int d\tau \int_{0}^{\pi} d \sigma
\left[ G_{ij} \partial_{\alpha} X^i \partial_{\beta} X^j \eta^{\alpha
    \beta} - B_{ij} \epsilon^{\alpha \beta} \partial_{\alpha} X^i
\partial_{\beta} X^j \right]  + S_{boundary}
\label{spluss}
\end{eqnarray}
where $S_{boundary}$ is equal to:
\[
S_{boundary} = - q_0 \int d \tau A_{i}^{(0)} \partial_{\tau} X^{i}
|_{\sigma =0} - q_{\pi} \int d \tau A_{i}^{(\pi)} \partial_{\tau} X^{i}
|_{\sigma =\pi} =
\]
\begin{eqnarray}
= \frac{q_0}{2} \int d \tau F_{ij}^{(0)} X^j {\dot{X}}^{i}|_{\sigma
=0} +\frac{q_{\pi}}{2} \int d \tau F_{ij}^{(\pi)} X^j
{\dot{X}}^{i}|_{\sigma =\pi}~~;~~~i, j =1 \dots {\hat{d}} \label{s2}
\end{eqnarray}
where $q_0$ and $q_{\pi}$ are the charges located at the two
end-points and, for the sake of generality, we keep $i$ and $j$ to vary
between $1$ and ${\hat{d}}$ (for a D9 brane ${\hat{d}}=6$).
We have used the expression $A_{i} = -
\frac{1}{2} F_{ij} X^j$ with a constant field strength.
From the previous action one can write the
equation of motion in the bulk given by:
\begin{eqnarray}
\partial_{\alpha} [ G_{ij} \partial^{\alpha} X^j ] =0 \Rightarrow
\partial_{\alpha} [  \partial^{\alpha} X^i ] =0
\label{bulkequa}
\end{eqnarray}
and the two boundary conditions at $\sigma =0, \pi$:
\begin{eqnarray}
\left[ G_{ij} \partial_{\sigma} X^j + ( B_{ij} - 2 \pi \alpha' q_0
  F_{ij}^{(0)}  )
  \partial_{\tau} X^j \right]_{\sigma =0} =0
\label{bou0}
\end{eqnarray}
and
\begin{eqnarray}
\left[ G_{ij} \partial_{\sigma} X^j + ( B_{ij} + 2 \pi \alpha' q_{\pi}
  F_{ij}^{(\pi)}  )
  \partial_{\tau} X^j \right]_{\sigma =\pi} =0~~.
\label{boupi}
\end{eqnarray}
The most general solution of the bulk equation in Eq. (\ref{bulkequa}) is given
by:
\begin{eqnarray}
X^{i} ( \sigma, \tau) = F^{i} (\tau + \sigma) + G^{i} ( \tau - \sigma )
\label{solu83}
\end{eqnarray}
with $F^{i} (\tau + \sigma)$  and $G^{i} ( \tau - \sigma )$ arbitrary
functions. By inserting Eq. (\ref{solu83}) in the boundary conditions in
Eq.s (\ref{bou0}) and (\ref{boupi}) we get:
\begin{eqnarray}
\partial_{\tau} G^{i} (\tau) = ( R_0 )^{i}_{\,\,j}
\partial_{\tau} F^{j} (\tau)~~;~~ \partial_{\tau} F^{i} (\tau + \pi) =
(R_{\pi}^{-1} R_0 )^{i}_{\,\,j} \partial_{\tau} F^{j} (\tau - \pi)
\label{boun31}
\end{eqnarray}
where
\begin{eqnarray}
(R_0 )^{i}_{\,\,j} = [(1 - {\cal{B}}_0 )^{-1} ( 1 + {\cal{B}}_0
)]^{i}_{\,\,j} ~~;~~
( R_{\pi} )^{i}_{\,\,j} = [(1 - {\cal{B}}_{\pi} )^{-1}
( 1 + {\cal{B}}_{\pi} )]^{i}_{\,\,j}
\label{R0pi}
\end{eqnarray}
and
\begin{eqnarray}
{\cal{B}}^{i}_{0\,\,j} = G^{ik} ( B_{kj} - 2 \pi \alpha' q_0 F_{kj}^{(0)})
~~;~~
{\cal{B}}^{i}_{\pi\,\,j} = G^{ik} ( B_{kj} + 2 \pi \alpha' q_{\pi}
F_{kj}^{(\pi)} )~~.
\label{def73}
\end{eqnarray}
Notice that the matrices $R_0$ and $R_{\pi}$ and therefore also
the product $R \equiv R_{\pi}^{-1} R_{0}$ are orthogonal matrices, i.e.
$
R_{(0, \pi)}^{T} = R_{(0, \pi)}^{-1}~,~R^T = R^{-1}.
$
An orthogonal matrix can always be brought into a diagonal form
with eigenvalues $\lambda$ such that $| \lambda | =1$. Moreover,
being orthogonal, such a matrix is unitary and real and therefore
in a space with an even number of dimensions (${\hat{d}}$ is even)
for each eigenvalue $\lambda$ it exists also an eigenvalue
$\lambda^{*}$.
Therefore one can always bring the orthogonal matrix $R$ in the
following form:
\begin{eqnarray}
R^{i}_{\,\,j} = {e}^{-2 \pi i \nu_{i}} \delta^{i}_{\,\,j}
\label{diago83}
\end{eqnarray}
where $\nu_{2a} = - \nu_{2a -1}$ for $a=1 \dots \frac{{\hat{d}}}{2}$.
$R$ is an orthogonal matrix of dimensionality ${\hat{d}}$  that  can be
diagonalized with eigenvectors and eigenvalues satisfying the
following equations:
\begin{eqnarray}
R^{i}_{\,\,j} C_{a}^{j} = {e}^{-2i \pi \nu_a} C_{a}^{i}~~;~~
R^{i}_{\,\,j} C_{a}^{*j} = {e}^{ 2i \pi \nu_a} C_{a}^{*i}~~;~~a=1
\dots \frac{{\hat{d}}}{2}
\label{eige54}
\end{eqnarray}
with~\footnote{In principle $\nu_a$ varies in the interval $0 \leq
  \nu_a < 1$ but we can restrict ourselves to the smaller interval
$0 \leq \nu_{a} \leq 1/2$ because of the freedom that we have in
  defining $C$ and $C^{*}$.  }
$0 \leq \nu_{a} \leq 1/2$. The eigenvectors $C$ and $C^{*}$
are orthonormal satisfying the
conditions:
\begin{eqnarray}
C_{a}^{*i} G_{ij} C_{b}^{j} = \delta_{ab}~~;~~C_{a}^{i} G_{ij} C_{b}^{j} =
C_{a}^{*i} G_{ij} C_{b}^{*j} = 0~~.
\label{norma67}
\end{eqnarray}
The quantities $\nu_{a}$ may be zero. This
happens when  $ \det ( q_0 F^{(0)} + q_{\pi} F^{(\pi)} )_{ij} =0$ and in this case
$C$ and $C^{*}$ may be taken real.
In the following we will assume that the previous matrix has nonzero entries
only along the directions  $ 1 \dots d$,  with $d \leq {\hat{d}}$ and even, and that the
determinant of its not null submatrix is different from zero, i.e.
$ \det (q_0 F^{(0)} + q_{\pi} F^{(\pi)} )_{AB} \neq 0$ for $0\leq
A,B \leq d$. All other entries are vanishing. This means
that $R$ has the form given in Eq. (\ref{diago83}) for $i,j =1 \dots
d$, while the remaining diagonal elements are equal to $1$
($\nu_{2a} =\nu_{2a-1}=0$ for $\frac{d}{2} < a \leq
\frac{{\hat{d}}}{2}$).

The equation that the boundary state must satisfy can be derived from
Eq. (\ref{bou0}) with the substitution $\sigma \leftrightarrow \tau$.
In so doing one gets:
\begin{eqnarray}
\left[ G_{ij} \partial_{\tau} X^j + ( B_{ij} - 2 \pi \alpha' q F_{ij} )
  \partial_{\sigma} X^j \right]_{\tau =0} | B \rangle =0~~.
\label{bou0bs}
\end{eqnarray}
Inserting in the previous equation the mode expansion for a closed
string:
\begin{eqnarray}
X^{i} (\tau, \sigma)& =  & x^i + \sqrt{\alpha'} \left[ 2 {\hat{m}}^i
\sigma + 2G^{ij}  \left(  {\hat{n}}_j -  B_{jk}
{\hat{m}}^k  \right)  \tau \right] + \nonumber \\
&&
+ i\frac{\sqrt{2 \alpha' }}{2}
\sum_{n \neq 0} \frac{1}{n}\left[\alpha_{n}^{i} e^{-2in(\tau -
    \sigma)}
+{\tilde{\alpha}}_{n}^{i} e^{-2in (\tau + \sigma)}  \right].
\label{mod48}
\end{eqnarray}
gives the following conditions:
\begin{eqnarray}
({\hat{n}}_i - 2 \pi \alpha' q F_{ij} {\hat{m}}^j ) | B \rangle =0
\label{zeromo}
\end{eqnarray}
and
\begin{eqnarray}
\left( {\cal{E}}_{ij} \alpha_{n}^{j} + {\cal{E}}^{T}_{ij}
{\tilde{\alpha}}^{j}_{-n} \right) |B \rangle =0~~;~~ {\cal{E}}_{ij} =
G_{ij} - B_{ij} + 2 \pi \alpha' q F_{ij}~~,
\label{othemo2}
\end{eqnarray}
being $qF=-q_\pi F_\pi$ on the boundary in $\sigma=\pi$ and
$qF=q_0F_0$  on the boundary in  $\sigma=0$.
In Eq.s  (\ref{mod48}) and (\ref{zeromo}) we have inserted the hat
to remember that ${\hat{n}}$ and ${\hat{m}}$ are operators.

The boundary state satisfying Eq.s (\ref{zeromo}) and
(\ref{othemo2}) is given by~\cite{IGOR}:
\[
| B \rangle_{0, \pi} = C_{0, \pi} N_{0, \pi} W_{0,\pi}
 \prod_{n=1}^{\infty} \left [ {e}^{- \frac{1}{n}
  \alpha^{i}_{-n} G_{ik} ({\cal{E}}^{-1})^{kh} ({\cal{E}}^{T})_{hj}
{\tilde{\alpha}}_{-n}^{j}} \right] \times
\]
\begin{eqnarray}
\times
\sum_{r_i , s^j \in \mathbb{Z}}  \delta_{
  n_i \mp 2 \pi \alpha' q_{0, \pi} F_{ij}^{(0, \pi)}  m^j }
|n_i =  \frac{r_i}{W_{0, \pi}^{i}} \rangle  | m^j = W^{j}_{0, \pi}
s^j \rangle   | 0_{\alpha ,
  {\tilde{\alpha}}}  \rangle
\label{bounda23}
\end{eqnarray}
where the plus sign refers to the boundary defined at $\sigma=\pi$
while the minus sign to the other one. $W_{0}^{i}~~ (W_{\pi}^{i})$ is
the wrapping number of the brane at $\sigma=0~~(\sigma= \pi)$ along
the $i$th direction and $W_{0, \pi} = \prod_{i=1}^{{\hat{d}}}
W^{i}_{0, \pi}$. Furthermore, $N_{0,\pi}$ is the number of D9 branes.

The gauge fields $q_{0,\pi} F_{ij}^{(0,\pi)}$ cannot be
arbitrary because the following quantity corresponding to the first
Chern class, given by:
\begin{eqnarray}
(c_1)_{ij} = \frac{1}{2 \pi} \int_{(i,j)} \frac{dx^i \wedge dx^j}{2} q_{0,\pi}
F_{ij}^{(0,\pi)} =
\frac{1}{2 \pi} {\cdot} (2 \pi \sqrt{\alpha'})^2  q_{0,\pi}
F_{ij}^{(0,\pi)} W^{i}_{0,\pi} W^{j}_{0,\pi} \equiv f_{ij}^{(0,\pi)},
\label{chern}
\end{eqnarray}
has to be an integer.
It is important to stress here that the coordinate of the brane along
the $i$th compact direction is assumed to vary in the interval
$(0, 2 \pi \sqrt{\alpha'} W^i)$.  $C_{0, \pi}$ is a normalization constant
that in general can be determined by computing the annulus diagram
both in the open and closed string channel and by comparing the two
results. In some particular cases this constant is known and will allow
us to compute the number of Landau levels by
performing a modular transformation on the annulus amplitude computed
in the closed string channel.
The states  corresponding to the zero modes are
normalized~\footnote{The origin of the factor
$(2 \pi \sqrt{\alpha '})^{1/2}$ can be traced back to Eq. (2,19) of
Ref.~\cite{NBPS} and Eq. (6.28) of Ref.~\cite{ANTOII} with $\Phi = 2
\pi \sqrt{\alpha'}$.} for any compact direction $i$ as follows:
\begin{eqnarray}
\langle n_i | (n ')_{i} \rangle  = (2 \pi \sqrt{\alpha '})^{1/2}
\delta_{n_i , (n ')_i }~~;~~
\langle m^i | (m ')^i  \rangle =  (2 \pi \sqrt{\alpha '})^{1/2}
\delta_{m^i , (m')^i }
\label{norma954}
\end{eqnarray}

\section{One-loop amplitudes}
\label{closed}

In this section we
use the previously constructed boundary state for computing
the boundary-boundary interaction. We need the closed string
propagator, taken to be equal to:
\begin{eqnarray}
D = \frac{\alpha' \pi}{2} \delta_{L_0 - {\tilde{L}}_0 , 0}
\int_{0}^{\infty} dt \,\, {e}^{- \pi  t (L_0 + {\tilde{L}}_0 )}
\label{d}
\end{eqnarray}
where
\begin{eqnarray}
L_0 + {\tilde{L}}_0  =   N + {\tilde{N}} + \frac{1}{2}
\left[G_{ij} {\hat{m}}^i {\hat{m}}^j
  + ( {\hat{n}}_i - B_{ik} {\hat{m}}^{k} ) G^{ij}
( {\hat{n}}_j - B_{jh} {\hat{m}}^{h} )  \right]
\label{L064}
\end{eqnarray}
and
\begin{eqnarray}
N = \sum_{n=1}^{\infty} G_{ij} \alpha_{-n}^{i} \alpha_{n}^{j}~~;~~
{\tilde{N}} = \sum_{n=1}^{\infty} G_{ij} {\tilde{\alpha}}_{-n}^{i}
{\tilde{\alpha}}_{n}^{j}~~.
\label{L063}
\end{eqnarray}
Here we have omitted to write the ghost contribution.

We are now ready to compute the annulus diagram that is given by
$\langle B, f^{(0)} | D | B, f^{(\pi)} \rangle$.
Taking into account the $\delta$-function in Eq. (\ref{d}) we can
trade ${\tilde{N}}$ with $N$.
The contribution of the zero modes can be easily computed
and one gets:
\[
\sum_{r_i, s^i, r_{i}' , {s^{j}} ' \in \mathbb{Z} }
\langle n_i= \frac{r_i}{W^{i}_{0}} | \langle m^j = W_{0}^{j} s^j | \delta_{n_i
  - 2 \pi \alpha' q_0 F_{ij }^{(0)} m^j  } \delta_{n_{i}'   + 2 \pi
  \alpha' q_{\pi} F_{ij}^{(\pi)} {m^{j}}' } \times
\]
\[
\times
{e}^{- \frac{\pi}{2} t \left[G_{ij} {\hat{m}}^i {\hat{m}}^j
  + ( {\hat{n}}_i - B_{ik} {\hat{m}}^{k} ) G^{ij}
( {\hat{n}}_j - B_{jh} {\hat{m}}^{h} )  \right] } | (n')_i =
\frac{r_{i} '}{W_{\pi}^{i} } \rangle | (m')^j = W_{\pi}^{j} {s^{j}}
' \rangle
\]
\[
= \sum_{s^i, {s^{j}} ' \in \mathbb{Z}} \langle \frac{f_{ij}^{(0)}
  s^j}{W_{0}^{i}}  |
\langle W_{0}^{j} s^j |
{e}^{- \frac{\pi}{2} t \left[G_{ij} {\hat{m}}^i {\hat{m}}^j
  + ( {\hat{n}}_i - B_{ik} {\hat{m}}^{k} ) G^{ij}
( {\hat{n}}_j - B_{jh} {\hat{m}}^{h} )  \right] } |-
\frac{f_{ij}^{(\pi)} {s^{j}} '}{W_{\pi}^{i}} \rangle | W_{\pi}^{j}
{s^j} ' \rangle
\]
\[
= (2 \pi \sqrt{\alpha'})^{\hat{d}}
\sum_{s^i, {s^{j}} ' \in \mathbb{Z}} \delta_{W_{0}^{j} s^j
  -W_{\pi}^{j} (s ')^j } \delta_{ f^{(0)}_{ij} s^j/ W_{0}^{i} +
f^{(\pi)}_{ij} (s ')^j / W_{\pi}^{i}} \times
\]
\begin{eqnarray}
\times
{e}^{- \frac{\pi}{2} t \left[G_{ij} W^{i}_{0} s^i W^{j}_{0} s^j
  + ( \frac{f^{(0)}_{ik} s^k}{W_{0}^{i}} -
B_{ik} W_{0}^{k} s^k ) G^{ij} ( \frac{f^{(0)}_{jh}
s^h}{W_{0}^{j}} - B_{jh} W_{0}^{h} s^h )  \right] }
\label{noze88}
\end{eqnarray}
where $f_{ij}^{(0,\pi)}$ is defined in Eq. (\ref{chern})
and we have used Eq.s (\ref{norma954}). In the case of the bosonic
string (superstring) ${\hat{d}} = 22~~(6)$.

It is easy to see that the first $\delta$-function can be satisfied only
if
\begin{eqnarray}
s^j = \frac{W^j_{lcm}}{W_{0}^{j}} u^{j}~~;~~(s ')^j  =
\frac{W^j_{lcm}}{W_{\pi}^{j}} u^{j} \label{esse}
\end{eqnarray}
where $u^j$ is an arbitrary integer and  $W^{i}_{lcm}$ is the
least common multiple of $W_{0}^{i}$ and $ W_{\pi}^{i}$.
By inserting the previous values in the other $\delta$-function one can
write it as $\delta_{ ( q_0 F_{ij}^{(0)} + q_{\pi} F_{ij}^{(\pi)}
)W^{j}_{lcm} u^j }$  getting
\begin{eqnarray}
(2 \pi \sqrt{\alpha'})^{\hat{d}} \sum_{u^j \in \mathbb{Z}} \delta_{(
q_0 F_{ij}^{(0)} + q_{\pi} F_{ij}^{(\pi)} )W^j_{lcm} u^j } {e}^{-
\frac{\pi}{2} t u^i
  W^i_{lcm} {\cal{G}}_{ij} u^j W^j_{lcm} }
\label{ze84}
\end{eqnarray}
in terms of the open string metric ${\cal{G}}_{ij}$ defined by:
\begin{eqnarray}
{\cal{G}}_{ij} \equiv  G_{ij} - {\cal B}_{ik} G^{kh} {\cal B}_{hj} =
{\cal{E}}^{T}_{ik} G^{kh} {\cal{E}}_{hj}~~;~~
{\cal{B}}_{ij} \equiv B_{ij} - 2 \pi \alpha' q_0
F_{ij}^{(0)}~~;~~{\cal{E}}_{ij}
\equiv G_{ij} - {\cal{B}}_{ij}~~.
\label{openme}
\end{eqnarray}
The contribution of the non-zero modes is given by:
\begin{eqnarray}
\prod_{n=1}^{\infty} \frac{1}{\det \left[ \delta^{i}_{\,\,j} -
    \left({\cal{E}}^{-1}_{0} {\cal{E}}^{T}_{0} \right)^{i}_{\,\,h}
G^{hk} \left( {\cal{E}}_{\pi}
    {\cal{E}}_{\pi}^{-1 T}    \right)_{kj} {e}^{- 2 \pi nt} \right]}
\label{noze84}
\end{eqnarray}
where the following commutation relations have been used:
\begin{eqnarray}
[ \alpha_{n}^{i} , \alpha_{m}^{j} ] = n \delta_{n+m;0} G^{ij}~~;~~
[ {\tilde{\alpha}}_{n}^{i} , {\tilde{\alpha}}_{m}^{j} ]
= n \delta_{n+m;0} G^{ij}~~.
\label{commu72}
\end{eqnarray}
Putting all factors together and adding the contribution of
the four non-compact directions and of the ghosts lead, in the case
of the bosonic string, to the following expression~\cite{IGOR}
($q \equiv {e}^{-\pi t}$):
\[
\langle B, f^{(0)} | D | B, f^{(\pi)} \rangle  =  C_0 N_0 W_0
C_{\pi} N_{\pi} W_{\pi}  (2 \pi \sqrt{\alpha'})^{\hat{d}}
\frac{\alpha' \pi}{2} V_4 \sum_{n^j \in \mathbb{Z}} \delta_{( q_0
F_{ij}^{(0)} + q_{\pi} F_{ij}^{(\pi)} )W^j_{lcm} u^j }
\int_{0}^{\infty} dt
 {\times}
\]
\begin{eqnarray}
 {\times}  {e}^{- \frac{\pi}{2} t u^i
  W^i_{lcm} {\cal{G}}_{ij} u^j W^j_{lcm} } {\times}
\frac{q^{- {\hat{d}}/12}}{(f_1 (q))^{2}}
\prod_{n=1}^{\infty} \frac{1}{\det
\left[ \delta^{i}_{\,\,j} -
    \left({\cal{E}}^{-1}_{0} {\cal{E}}^{T}_{0} \right)^{i}_{\,\,h}
    G^{hk}  \left( {\cal{E}}_{\pi}
    {\cal{E}}_{\pi}^{-1 T}    \right)_{kj} {e}^{-2 \pi nt} \right]}~~.
\label{annuclo}
\end{eqnarray}
Eq. (\ref{annuclo}) can be easily generalized to the case of superstring
becoming:
\[
\langle B, f^{(0)} | D | B, f^{(\pi)} \rangle  =  C_0 C_{\pi} (2 \pi
\sqrt{\alpha'})^{6}  \frac{\alpha' \pi}{2} V_4 \sum_{u^j \in
\mathbb{Z}} \delta_{( q_0 F_{ij}^{(0)} + q_{\pi} F_{ij}^{(\pi)}
)W^j_{lcm} u^j }
\times
\]
\[
N_0 N_{\pi} W_0 W_{\pi}  \int_{0}^{\infty} dt {e}^{- \frac{\pi}{2} t
u^i
  W^i_{lcm} {\cal{G}}_{ij} u^j W^j_{lcm} } {\times}
\]
\[
 \frac{1}{2}\left\{ \frac{1}{q} \left[ \prod_{n=1}^{\infty}
\frac{ \det \left[ \delta^{i}_{\,\,j} +
    \left({\cal{E}}^{-1}_{0} {\cal{E}}^{T}_{0} \right)^{i}_{\,\,h}
    G^{hk}  \left( {\cal{E}}_{\pi}
    {\cal{E}}_{\pi}^{-1 T}
\right)_{kj} q^{2n-1} \right] (1 + q^{2n-1})^2 }{\det
\left[ \delta^{i}_{\,\,j} -
    \left({\cal{E}}^{-1}_{0} {\cal{E}}^{T}_{0} \right)^{i}_{\,\,h}
    G^{hk}  \left( {\cal{E}}_{\pi}
    {\cal{E}}_{\pi}^{-1 T}    \right)_{kj} q^{2n} \right] (1 -
q^{2n})^2 } +  \right.
\right.
\]
\[
- \left. \prod_{n=1}^{\infty}
\frac{ \det \left[ \delta^{i}_{\,\,j} -
    \left({\cal{E}}^{-1}_{0} {\cal{E}}^{T}_{0} \right)^{i}_{\,\,h}
    G^{hk}  \left( {\cal{E}}_{\pi}
    {\cal{E}}_{\pi}^{-1 T}    \right)_{kj} q^{2n-1} \right]
(1 - q^{2n-1})^2}{\det
\left[ \delta^{i}_{\,\,j} -
    \left({\cal{E}}^{-1}_{0} {\cal{E}}^{T}_{0} \right)^{i}_{\,\,h}
    G^{hk}  \left( {\cal{E}}_{\pi}
    {\cal{E}}_{\pi}^{-1 T}    \right)_{kj} q^{2n} \right](1 -
q^{2n})^2 } \right] +
\]
\begin{eqnarray}
\left. - \left[ 2^{4-\hat{d}/2} \prod_{a=1}^{\hat{d}/2} (2 \cos \pi \nu_a
)\right] \prod_{n=1}^{\infty} \frac{ \det \left[ \delta^{i}_{\,\,j} +
    \left({\cal{E}}^{-1}_{0} {\cal{E}}^{T}_{0} \right)^{i}_{\,\,h}
    G^{hk}  \left( {\cal{E}}_{\pi}
    {\cal{E}}_{\pi}^{-1 T}    \right)_{kj} q^{2n} \right](1 +
q^{2n})^2 }{\det
\left[ \delta^{i}_{\,\,j} -
    \left({\cal{E}}^{-1}_{0} {\cal{E}}^{T}_{0} \right)^{i}_{\,\,h}
    G^{hk}  \left( {\cal{E}}_{\pi}
    {\cal{E}}_{\pi}^{-1 T}    \right)_{kj} q^{2n} \right](1 -
q^{2n})^2 } \right\}~~.
\label{annuclosu}
\end{eqnarray}
By using Eq. (\ref{R0pi}) together with the following equations:
\begin{eqnarray}
(1 - {\cal{B}}_{0} )^{i}_{\,\,j} = G^{ik} {\cal{E}}^{(0)}_{kj}
   \equiv ({\cal{E}}_{0})^{i}_{\,\,j}~~;~~
(1 - {\cal{B}}_{\pi} )^{i}_{\,\,j} = G^{ik} {\cal{E}}^{(\pi)}_{kj}
   \equiv ({\cal{E}}_{\pi})^{i}_{\,\,j}
\label{inte41}
\end{eqnarray}
and the fact that under a determinant we can change the order of the two
matrices
in the last line of Eq. (\ref{annuclo}), we can rewrite Eq. (\ref{annuclosu})
as follows:
\[
\langle B, f^{(0)} | D | B, f^{(\pi)} \rangle  =  C_0 C_{\pi} (2 \pi
\sqrt{\alpha'})^{6}  \frac{\alpha' \pi}{2} V_4 \sum_{u^j \in
\mathbb{Z}} \delta_{( q_0 F_{ij}^{(0)} + q_{\pi} F_{ij}^{(\pi)}
)W^j_{lcm} u^j } \times
\]
\[
\times N_0 N_{\pi} W_0 W_{\pi} \int_{0}^{\infty} dt \,\, {e}^{-
\frac{\pi}{2} t u^i
  W^i_{lcm} {\cal{G}}_{ij} u^j W^j_{lcm} } {\times}
\]
\[
\frac{1}{2} \left\{
\frac{1}{q} \left[\prod_{n=1}^{\infty} \frac{ \det \left[ \delta^{i}_{\,\,j} +
    R^{i}_{\,\,j}  q^{2n-1} \right](1 + q^{2n+1})^2 }{\det
\left[ \delta^{i}_{\,\,j} -
    R^{i}_{\,\,j}  q^{2n} \right](1 - q^{2n})^2 } -
\prod_{n=1}^{\infty} \frac{ \det \left[ \delta^{i}_{\,\,j} -
    R^{i}_{\,\,j}  q^{2n-1} \right](1 - q^{2n+1})^2 }{\det
\left[ \delta^{i}_{\,\,j} -
    R^{i}_{\,\,j}  q^{2n} \right](1 - q^{2n})^2 } \right] + \right.
\]
\begin{eqnarray}
\left. - \left[ 2^{4-\hat{d}/2} \prod_{a=1}^{\hat{d}/2} (2 \cos \pi \nu_a
)\right] \prod_{n=1}^{\infty} \frac{ \det \left[ \delta^{i}_{\,\,j} +
    R^{i}_{\,\,j}  q^{2n} \right](1 - q^{2n+1})^2 }{\det
\left[ \delta^{i}_{\,\,j} -
    R^{i}_{\,\,j}  q^{2n} \right](1 - q^{2n})^2 } \right\}
\label{annuclo21}
\end{eqnarray}
where $R = R_{\pi}^{-1} R_0$ and the matrices $R_0$ and $R_{\pi}$ are
the ones introduced  in Eq. (\ref{R0pi}).
The last two lines of the previous equation can be written as follows~\footnote{We use the definition of the $\Theta$-functions given in App. A of Ref \cite{DLMP} where References to previous papers dealing with strings interacting with gauge fields with constant field strength, can also be found.}:
\[
\frac{1}{2}\left\{ \prod_{a=1}^{d/2} \left[
\frac{\Theta_{3} ( \nu_a | it)}{\Theta_{1} ( \nu_a | it)} \right]
\left( \frac{ f_3 (q)}{f_1 (q)} \right)^{8 -d} -
\prod_{a=1}^{d/2} \left[
\frac{\Theta_{4} ( \nu_a | it)}{\Theta_{1} ( \nu_a | it)} \right]
\left( \frac{ f_4 (q)}{f_1 (q)} \right)^{8 -d} + \right.
\]
\begin{eqnarray}
\left. -\prod_{a=1}^{d/2} \left[
\frac{\Theta_{2} ( \nu_a | it)}{\Theta_{1} ( \nu_a | it)} \right]
\left( \frac{ f_2 (q)}{f_1 (q)} \right)^{8 -d}
\right\}\prod_{a=1}^{d/2} ( -2 \sin \pi \nu_a )~~.
\label{annuclo32}
\end{eqnarray}
This implies that Eq. (\ref{annuclo21}) becomes:
\[
\langle B, f^{(0)} | D | B, f^{(\pi)} \rangle  =  C_0 C_{\pi} (2 \pi
\sqrt{\alpha'})^{6}  \frac{\alpha' \pi}{2} V_4 \sum_{u^j \in
\mathbb{Z}} \delta_{( q_0 F_{ij}^{(0)} + q_{\pi} F_{ij}^{(\pi)}
)W^j_{lcm} u^j } \times
\]
\[
\times N_0 N_{\pi} W_0 W_{\pi} \int_{0}^{\infty} dt \,\, \sum_{u^i,
u^j} {e}^{- \frac{\pi}{2} t u^i
  W^i_{lcm} {\cal{G}}_{ij} u^j W^j_{lcm} }
\prod_{a=1}^{d/2} ( -2 \sin \pi \nu_a ) {\times}
\]
\[
\frac{1}{2}\left\{ \prod_{a=1}^{d/2} \left[
\frac{\Theta_{3} ( \nu_a | it)}{\Theta_{1} ( \nu_a | it)} \right]
\left( \frac{ f_3 (q)}{f_1 (q)} \right)^{8 -d} -
\prod_{a=1}^{d/2} \left[
\frac{\Theta_{4} ( \nu_a | it)}{\Theta_{1} ( \nu_a | it)} \right]
\left( \frac{ f_4 (q)}{f_1 (q)} \right)^{8 -d} + \right.
\]
\begin{eqnarray}
\left. -\prod_{a=1}^{d/2} \left[
\frac{\Theta_{2} ( \nu_a | it)}{\Theta_{1} ( \nu_a | it)} \right]
\left( \frac{ f_2 (q)}{f_1 (q)} \right)^{8 -d}
\right\} ~~.
\label{fina67}
\end{eqnarray}
It is straightforward to compute Eq. (\ref{fina67}) for
$\hat{d}=d = 6$ and $d=0$. In the first case we can use the
following equation:
\begin{eqnarray}
1-R \equiv 1 - R_{\pi}^{-1} R_0 = 1 - (1+ {\cal{B}}_{\pi} )^{-1} (1-
 {\cal{B}}_{\pi} ) (1- {\cal{B}}_{0} )^{-1} (1+ {\cal{B}}_{0} ) & = &
 \nonumber \\
= (1+ {\cal{B}}_{\pi} )^{-1}  \left[(1+ {\cal{B}}_{\pi} ) (1-
 {\cal{B}}_{0} ) - (1- {\cal{B}}_{\pi} )(1+ {\cal{B}}_{0} )
\right] (1- {\cal{B}}_{0} )^{-1} &=& \nonumber \\
= (1+ {\cal{B}}_{\pi} )^{-1} G^{-1} (4 \pi \alpha' ) ( q_{\pi} F_{\pi} + q_0
 F_{0} ) (1- {\cal{B}}_{0} )^{-1}
\label{equa81}
\end{eqnarray}
together with
\begin{eqnarray}
\sqrt{\det ( 1 - R_{\pi}^{-1} R_0 )  } =
\prod_{a=1}^{{\hat{d}}/2} ( 2 \sin \pi \nu_a )
\label{equa29}
\end{eqnarray}
in order to rewrite ($0\leq\nu_a \leq 1/2$)
\begin{eqnarray}
\prod_{a=1}^{{\hat{d}}/2} ( -2 \sin \pi \nu_a ) =
\frac{(-1)^{{\hat{d}}/2} \sqrt{ \det G_{ij}} \sqrt{{\rm det}[4 \pi
\alpha' ( q_0 F^{(0)} + q_{\pi}
  F^{(\pi)})_{ij}] }  }{\sqrt{\det( G_{ij} + B_{ij} + 2 \pi \alpha'
  q_{\pi} F^{(\pi)}_{ij}) }\sqrt{\det( G_{ij} + B_{ij} - 2 \pi \alpha'
  q_{0} F^{(0)}_{ij} }) }~~.
\label{eq673}
\end{eqnarray}
It is easy to convince oneself that, for $d={\hat{d}}= 6$, one gets:
\begin{eqnarray}
C_{0} = \frac{T_9}{2} \sqrt{\det( G_{ij} + B_{ij} - 2 \pi
\alpha'
  q_{0} F^{(0)}_{ij}) }/( \det G_{ij})^{1/4}~~;~~
T_9 = \frac{\sqrt{\pi}}{(2 \pi \sqrt{\alpha'})^{6}}
\label{C0}
\end{eqnarray}
and
\begin{eqnarray}
C_{\pi} = \frac{T_9}{2} \sqrt{\det( G_{ij} + B_{ij} + 2 \pi
\alpha'
  q_{\pi} F^{(\pi)}_{ij}) }/( \det G_{ij})^{1/4}~~.
\label{Cpi}
\end{eqnarray}
In fact the first two factors in the two previous equations are
precisely those that one also gets  in non-compact
space~\cite{ANTOI,ANTOII}. The last factor in the denominator is
instead  peculiar of a compact space and is already present in the
case of a single direction compactified on a circle of radius $R$ as
one can immediately check. After inserting Eq.s (\ref{C0}) and (\ref{Cpi})
in Eq. (\ref{fina67}) we get (${\hat{d}}=6$):
\[
\langle B, f^{(0)} | D | B, f^{(\pi)} \rangle^{d=6} = -
\frac{V_4 N_0 N_{\pi}}{(8 \pi^2
  \alpha')^2}  N_{LL} \int_{0}^{\infty} dt  \frac{1}{2}
  \left\{ \prod_{a=1}^{3}
\frac{\Theta_{3} ( \nu_a | it)}{\Theta_{1} ( \nu_a | it)}
\left( \frac{ f_3 (q)}{f_1 (q)} \right)^{2} +  \right.
\]
\begin{eqnarray}
\left. - \prod_{a=1}^{3} \left[
\frac{\Theta_{4} ( \nu_a | it)}{\Theta_{1} ( \nu_a | it)} \right]
\left( \frac{ f_4 (q)}{f_1 (q)} \right)^{2} -  \prod_{a=1}^{3} \left[
\frac{\Theta_{2} ( \nu_a | it)}{\Theta_{1} ( \nu_a | it)} \right]
\left( \frac{ f_2 (q)}{f_1 (q)} \right)^{2}   \right\}
\label{fina68}
\end{eqnarray}
where $N_{LL}$ is the number of Landau levels, given by\cite{BT2}:
\begin{eqnarray}
N_{LL}= (2\pi\sqrt{\alpha'})^6 W_{0} W_{\pi} \sqrt{{\rm det}
\left(\frac{q_\pi F^{(\pi)} +q_0 F^{(0)}}{2\pi}\right)}
\label{llev}
\end{eqnarray}
and we have used the relation:
\begin{eqnarray}
\left( \frac{T_9}{2} \right)^2 ( 2 \pi \sqrt{\alpha'} )^6 \frac{
\alpha' \pi}{2} = \frac{1}{8  (8 \pi^2 \alpha')^2}~~. \label{rel92}
\end{eqnarray}
Eq. (\ref{fina68}) becomes in the open string channel ( $\tau = 1/t, k
={e}^{- \pi \tau}$):
\[
\langle B, f^{(0)} | D | B, f^{(\pi)} \rangle^{d=6}= \frac{i V_4 N_0
N_{\pi}N_{LL}}{(8\pi^2\alpha')^2} \int_0^{\infty}
\frac{d\tau}{\tau^3} \frac{1}{2}\!
\left[\left(\frac{f_3(k)}{f_1(k)}\right)^{2}
\prod_{a=1}^{3}\frac{\Theta_3\left(i
\nu_a\tau|i\tau\right)}{\Theta_1\left(i\nu_a\tau|i\tau\right)} +
\right.
\]
\begin{eqnarray}
\left. - \left(\frac{f_4(k)}{f_1(k)}\right)^{2}
\prod_{a=1}^{3}\frac{\Theta_4\left(i\nu_a\tau|i\tau\right)}{\Theta_1\left(i\nu_a\tau|i\tau\right)}
-\left(\frac{f_2(k)}{f_1(k)}\right)^{2}
\prod_{a=1}^{3}\frac{\Theta_2\left(i
\nu_a\tau|i\tau\right)}{\Theta_1\left(i\nu_a\tau|i\tau\right)}
\right]~~. \label{dyo}
\end{eqnarray}
Also the case $d=0$ is easy to treat. In fact in this case, by using Eq.s
(\ref{C0}) and (\ref{Cpi}) (that are still valid) with
$ (q_{0} F^{(0)}_{ij} + q_{\pi} F^{(\pi)})_{ij} =0$, it is easy to show
that Eq. (\ref{fina67}) reads as:
\[
\langle B, f^{(0)} | D | B, f^{(\pi)} \rangle^{d=0} =
\frac{V_4 N_0 N_{\pi} W_{0} W_{\pi}}{(8 \pi^2
  \alpha')^2}  \int_{0}^{\infty} dt \,\, \sum_{u^i, u^j}
{e}^{- \frac{\pi}{2} t u^i
  W^i_{mcm} {\cal{G}}_{ij} u^j W^j_{mcm} } \times
\]
\begin{eqnarray}
\left[ \det \left( \frac{{\cal{G}}_{ij}}{2}\right) \right]^{1/2}\frac{1}{2}
\left[ \left( \frac{ f_3 (q)}{f_1 (q)} \right)^{8}
-  \left( \frac{ f_4 (q)}{f_1 (q)} \right)^{8} -
\left( \frac{ f_2 (q)}{f_1 (q)} \right)^{8} \right]
\label{d=0}
\end{eqnarray}
where we have used the relation:
\begin{eqnarray}
\left[ \det \left( \frac{{\cal{G}}_{ij}}{2}\right) \right]^{1/2} =
\frac{\det ( G+ B - 2 \pi \alpha' q_0 F^{(0)})_{ij} }{8 \sqrt{\det G_{ij}}}~~.
\label{rel34}
\end{eqnarray}
In the open string channel Eq. (\ref{d=0}) becomes:
\[
\langle B, f^{(0)} | D | B, f^{(\pi)} \rangle^{d=0} =
\frac{V_4 N_0 N_{\pi} }{(8 \pi^2
  \alpha')^2} W_{GCD}  \int_{0}^{\infty} \frac{d\tau}{\tau^3}
\sum_{u\in \mathbb{Z}^{6-d}}e^{-2\pi\tau\frac{u_i}{W^i_{lcm}}{\cal
G}^{ ij}\frac{u_j}{W^j_{lcm}}} \times
\]
\begin{eqnarray}
\times \frac{1}{2}\!
\left[\left(\frac{f_3(k)}{f_1(k)}\right)^{8}
-\left(\frac{f_4(k)}{f_1(k)}\right)^{8}
-\left(\frac{f_2(k)}{f_1(k)}\right)^{8} \right]
\label{dyo2}
\end{eqnarray}
where $W^{i}_{GCD}$ is the greatest
common divisor of $W_{0}^{i}$ and $ W_{\pi}^{i}$~\footnote{Remember that
$W_0 W_{\pi} = W_{lcm} W_{GCD}$.}.

The intermediate case is more difficult to treat and we will not
derive it in detail, but we will only give the final expression:
\[
\langle B, f^{(0)} | D | B, f^{(\pi)} \rangle = \frac{ V_4 N_0
N_{\pi} }{(8\pi^2\alpha')^2} W_{GCD}^{>d}  \int_0^{\infty} dt\,N_{LL}^{(d)}
(-)^{d/2} \left[{\rm det}\left(\frac{W_{lcm}^i{\cal
G}_{ij}^{>d}W_{lcm}^j}{2}\right)\right]^{1/2} \times
\]
\[
\sum_{u\in \mathbb{Z}^{6-d}}e^{-\frac{\pi\,t}{2}W^{i}_{lcm}u^i{\cal
G}_{ij}^{>d} u^jW^{j}_{lcm}}
\frac{1}{2}\!\left[\left(\frac{f_3(q)}{f_1(q)}\right)^{8-d}
\prod_{a=1}^{d/2}\frac{\Theta_3\left(\nu_a|i\,t\right)}{\Theta_1\left(\nu_a|i\,t\right)}
+ \right.
\]
\begin{eqnarray}
\left.
-\left(\frac{f_4(q)}{f_1(q)}\right)^{8-d}\prod_{a=1}^{d/2}\frac{
\Theta_4\left(\nu_a|i\,t\right)}{\Theta_1\left(\nu_a|i\,t\right)}
-\left(\frac{f_2(q)}{f_1(q)}\right)^{8-d}\prod_{a=1}^{d/2}\frac{
\Theta_2\left(\nu_a|i\,t\right)}{\Theta_1\left(\nu_a|i\,t\right)}
\right] \label{dyoc}
\end{eqnarray}
that becomes in the open string channel:
\begin{eqnarray}
\langle B, f^{(0)} | D | B, f^{(\pi)} \rangle
\!\!\!&=&\!\!\!\frac{V_4 N_0 N_{\pi}}{(8\pi^2\alpha')^2}
W_{GCD}^{>d} \int_0^{\infty}
\frac{d\tau}{\tau^3}N_{LL}^{(d)}(-i)^{d/2}\sum_{u\in
\mathbb{Z}^{6-d}}e^{-2\pi\tau\frac{u_i}{W^i_{lcm}}{\cal
G}^{>d ij}\frac{u_j}{W^j_{lcm}}}\nonumber\\
\!\!\!&{\times}&\!\!\!\!\!\frac{1}{2}\!
\left[\left(\frac{f_3(k)}{f_1(k)}\right)^{8-d}
\prod_{a=1}^{d/2}\frac{\Theta_3\left(i
\nu_a\tau|i\tau\right)}{\Theta_1\left(i\nu_a\tau|i\tau\right)}
-\left(\frac{f_4(k)}{f_1(k)}\right)^{8-d}
\prod_{a=1}^{d/2}\frac{\Theta_4\left(i\nu_a\tau|i\tau\right)}{\Theta_1\left(i\nu_a\tau|i\tau\right)}
\right.\nonumber\\
&-&\left.\left(\frac{f_2(k)}{f_1(k)}\right)^{8-d}
\prod_{a=1}^{d/2}\frac{\Theta_2\left(i
\nu_a\tau|i\tau\right)}{\Theta_1\left(i\nu_a\tau|i\tau\right)}
\right]\nonumber\\
\label{dyo3}
\end{eqnarray}
where by the upper index $(>d)$ we mean that the indices $i,j$ run in the
interval $d < i,j \leq {\hat{d}}$ and now the number of Landau levels
is given by:
\begin{eqnarray}
N_{LL}^{(d)}= (2\pi\sqrt{\alpha'})^{d} W_{0}^{d} W_{\pi}^{d} \sqrt{{\rm det}
\left(\frac{(q_\pi F^{(\pi)}+q_0F^{(0)})_{ij}}{2\pi}\right)^{(d)}}
\label{llev2}
\end{eqnarray}
with $W_{0, \pi}^{d} \equiv \prod_{i=1}^{d} W^{i}_{0, \pi}$. The upper
index $d$ of the matrix in the last term
indicates that we must compute the determinant of the $d \times d$
submatrix  whose entries are nonvanishing. Eq. (\ref{llev2}) generalizes to the torus $T^d$ the result obtained in Ref.~\cite{ACNY} for the torus $T2$.

In the last part of this paper, by comparing Eq.s (\ref{fina67}) and
(\ref{dyoc}), we are going to determine the normalization of the boundary
state in the general case assuming, as we have already done,
 that the matrix ${\rm det}(q_0 F^{(0)} +
q_{\pi} F^{(\pi)})_{AB} \neq 0$ only for $1 \leq  A,B \leq d$.
By comparing Eq.s (\ref{fina67}) and (\ref{dyoc}) we get:
\begin{eqnarray}
2^{d/2} N_{LL}^{(d)} \left(\det {\cal{G}}_{ij}^{(>d)} \right)^{1/2}
= W_{0}^{d} W_{\pi}^{d} {\hat{C}}_0 {\hat{C}}_{\pi}
\prod_{a=1}^{d/2} (2 \sin \pi\nu_a )~~;~~ C_{0,\pi} = \frac{T_9}{2}
{\hat{C}}_{0, \pi}~~. \label{nor59}
\end{eqnarray}
In order to fix ${\hat{C}}_{0, \pi}$ we need to use again Eq.
(\ref{equa81}), but in this general case $\det (1-R) =0$. In order
to get a nonvanishing result we have to restrict ourselves to the
determinant of the submatrix living in  the subspace of the
eigenvectors of $1-R$ with nonvanishing eigenvalues. This can be
done by grouping the eigenvectors with nonvanishing eigenvalues into
a $\hat d \times d$-dimensional matrix $L^\dagger$ and its hermitian
$L$:
\begin{eqnarray}
L_{\alpha}^{\,\,\,\,i} = \left( \begin{array}{c} C_{1}^{Ti} \\
                                         C_{1}^{\dagger i} \\
                                         \dots \\
                                         \dots \\
                                         C_{d/2}^{Ti} \\
                                          C_{d/2}^{\dagger i} \end{array}
\right)~~;~~L^{\dagger i}_{\,\,\,\,\alpha} = \left( \begin{array}{cccccc}
  C_{1}^{*i} &  C_{1}^{i} & \dots & \dots &  C_{d/2}^{*i} &
                                         C_{d/2}^{i} \end{array} \right)
\label{vec98}
\end{eqnarray}
with $\alpha=1...d$  and by computing the determinant of the
following $d \times d$ matrix:
\begin{eqnarray}
\sqrt{\det \left( L_{\alpha}^{\,\,\,i} G_{ik} (1 -R)^{k}_{\,\,j} L^{\dagger
    j}_{\,\,\,\,\beta} \right)} = \prod_{a=1}^{d/2} ( 2 \sin \pi\nu_a )~~.
\label{det56}
\end{eqnarray}
On the other hand, by using Eq. (\ref{equa81}) one can see that the
previous determinant is also equal to:
\[
\sqrt{ {\rm \det} \left( L_{\alpha}^{\,\,\,i} G_{ik} [(1+
  {\cal{B}}_{\pi})^{-1}]^{k}_{\,\,\, h} G^{hA} \right)}
\sqrt{ {\rm \det} \left[ 4 \pi \alpha' (q_0 F^{(0)} + q_{\pi}
F^{(\pi)} )_{AB} \right] } \times
\]
\begin{eqnarray}
\times \sqrt{{\rm \det} \left(  [ (1- {\cal{B}}_{0})^{-1} ]^{B}_{\,\,\,j}
L^{\dagger j}_{\,\,\,\,\beta}\right) }~~.
\label{det52}
\end{eqnarray}
By inserting this equation in Eq. (\ref{nor59}) and using
Eq. (\ref{llev2}), we get:
\begin{eqnarray}
C_0 = \frac{T_9}{2} \frac{[ \det {\cal{G}}^{(>d)}_{ij} ]^{1/4}
}{\sqrt{| \det \left( [ (1- {\cal{B}}_{0})^{-1} ]^{B}_{\,\,\,j} L^{\dagger
      j}_{\,\,\,\beta} \right)| }}~~;~~
C_{\pi} = \frac{T_9}{2} \frac{[ \det {\cal{G}}^{(>d)}_{ij} ]^{1/4}
}{\sqrt{| \det (L_{\alpha}^{\,\,\,i} G_{ik}
[ (1 + {\cal{B}}_{\pi} )^{-1} ]^{k}_{\,\,j} G^{jA}) | }}
\label{norma68}
\end{eqnarray}
where we have introduced the absolute value because Eq. (\ref{det52})
does not depend on the phases.
For $d=0$ the two denominators are absent and the two normalization constants
reduce to Eq.s (\ref{C0}) and (\ref{Cpi}) that are also valid for
$d=0$.
For $d= {\hat{d}}= 6$ the numerator is absent and in the denominator both the
indices $\alpha$ and $i$ run in the same interval. It  follows:
\begin{eqnarray}
C_0= \frac{T_9}{2} \frac{\sqrt{{\rm det}(1 -{\cal B}_0)}}{\sqrt{|{\rm \det}
  L | }} =\frac{T_9}{2} \sqrt{{\rm det}(1 -{\cal B}_0)} \left({\rm
  \det} G_{ij} \right)^{1/4}= \frac{T_9}{2}
\frac{\sqrt{{\rm det}(G-{\cal
B}_0)_{ij}}}{\left( {\rm \det} G_{ij}\right)^{1/4}}
\label{nc}
\end{eqnarray}
where we have used the relation $L_{\alpha}^{\,\,\,i} G_{ij} L^{\dagger
  j}_{\,\,\,\,\beta} = \delta_{\alpha \beta}$ that summarizes the ones
  in  Eq.s (\ref{norma67}). Furthermore Eq.s (\ref{C0}) and (\ref{nc})
are, as expected, in agreement.

We conclude by extending the previous calculation to Type I string
theory. In this case all the amplitudes must be divided by a factor
two due to the orientifold projection, and we have also to take into
account the interaction boundary-crosscap  given by:
\begin{eqnarray}
{\cal M}&=&-\frac{2^{5 -d/2} V_4}{(8\pi^2\alpha')^2}(-1)^{d/2}W^{>d} N[{\rm
det} \frac{{\cal G}_{ij}^{>d}}{2}]^{1/2}\int_0^\infty dt N_{LL}^{(d)}
\sum_{u^j\in\mathbb{Z}} e^{-2\pi t u^iW^i{\cal G}_{ij}^{>d}u^j
W^j}\nonumber\\
&\times&\frac{1}{2}\left[\prod_{a=1}^{d/2} \frac{\Theta_3 ({\hat{\nu}}_a
 |it+\frac{1}{2}) }{\Theta_{1} ( {\hat{\nu}}_a | it+\frac{1}{2})}
\left(\frac{f_3(iq)}{f_1(iq)} \right)^{8 -d}-\prod_{a=1}^{d/2}
\frac{ \Theta_4 ({\hat{\nu}}_a
 |it+\frac{1}{2}) }{\Theta_{1} ( {\hat{\nu}}_a | it+\frac{1}{2})}
 \left(\frac{f_4(iq)}{f_1(iq)}
  \right)^{8 -d}\right.\nonumber\\
   &&\left.-\prod_{a=1}^{d/2} \frac{\Theta_2 ( {\hat{\nu}}_a
  |it+\frac{1}{2}) }{\Theta_{1} ( {\hat{\nu}}_a | it+\frac{1}{2})}
 \left(\frac{f_2(iq)}{f_1(iq)} \right)^{8
  -d}\right]
 \label{m1}
 \end{eqnarray}
 where
 \begin{eqnarray}
 N_{LL}^{(d)}=W^d\,(2\pi\sqrt{\alpha'})^d \sqrt{ {\rm det} \left(
 \frac{2q\,F}{2\pi}\right)_{ij}} \label{nllm}
 \end{eqnarray}
and ${\hat{\nu}}_a$ are the eigenvalues of the matrix $R$ taken with $F_\pi=0$.

\section*{Acknowledgments}
This work has been partially supported by the European
Community's Human Potential Programme under contract
MRTN-CT-2004-005104 `Constituents, fundamental forces and symmetries
of the universe'.

\end{document}